\documentclass[letterpaper, 10 pt, conference]{ieeeconf}
\IEEEoverridecommandlockouts
\usepackage{graphicx} 
\usepackage{cite}
\usepackage{amssymb}
\usepackage{amsmath}
\usepackage{enumerate}
\usepackage{accents}
\usepackage{hyperref}
\hypersetup{
    hidelinks,
    pdftitle={Information-theoretic receding-horizon active learning of nonlinear dynamical systems},
    pdfauthor={Juncal Arbelaiz, Anushri Arora, Jonathan W. Pillow}
}
\usepackage{algorithm}
\usepackage{algorithmic}
\usepackage[dvipsnames]{xcolor}

\usepackage{caption}
\newcommand{\E}{\mathbb{E}}

\newcommand{\trp}{^\top} 

\newcommand{\Nrm}{\mathcal{N}}

\newcommand{\vx}{\mathbf{x}}

\newcommand{\va}{\mathbf{a}}

\newcommand{\vw}{\mathbf{w}}
\newcommand{\vy}{\mathbf{y}}
\newcommand{\vz}{\mathbf{z}}

\newcommand{\RR}{\mathbb{R}}
\newcommand{\mX}{\mathcal{X}}
\newcommand{\vphi}{\boldsymbol{\phi}}
\newcommand{\vmu}{\boldsymbol{\mu}}

\newtheorem{ass}{Assumption}[section]

\let\OLDthebibliography\thebibliography
\renewcommand\thebibliography[1]{%
  \OLDthebibliography{#1}%
  \setlength{\itemsep}{-0.3ex}%
  \setlength{\parskip}{0pt}%
  \setlength{\parsep}{0pt}%
}

\title{\LARGE \bf
Information-theoretic receding-horizon active learning of nonlinear dynamical systems
}
\author{Juncal Arbelaiz$^{a}$, Anushri Arora$^{b}$, 
       Jonathan W. Pillow$^{a}$
    \thanks{}
    \thanks{JA acknowledges the support of a C.V. Starr postdoctoral fellowship and the Burroughs Wellcome Fund Career Award at the Scientific Interface (ID 1369192). JWP was supported by grants from the Simons Collaboration on the Global Brain (SCGB AWD543027), the NIH BRAIN initiative (9R01DA056404-04), a U19 NIH-NINDS BRAIN Initiative Award (U19NS104648, U19NS123716).}
    \thanks{$^{a}$Princeton Neuroscience Institute and Center for Statistics and Machine Learning, Princeton University; \texttt{\{arbelaiz,jpillow\}@princeton.edu}}
    \thanks{$^{b}$Computer Science Department, Princeton University; \texttt{aa1698@princeton.edu}}
    \thanks{© 2026 IEEE. Personal use of this material is permitted. Permission from IEEE must be obtained for all other uses, in any current or future media, including reprinting/republishing this material for advertising or promotional purposes, creating new collective works, for resale or redistribution to servers or lists, or reuse of any copyrighted component of this work in other works.}
}

\begin{document}

\maketitle
\begin{abstract}
Accurately learning nonlinear dynamics from a finite-duration experiment requires the efficient collection of informative data. We address this challenge for stochastic controlled nonlinear dynamical systems whose state is observed along a single trajectory. Our goal is to reconstruct the unknown controlled state-increment map over a prescribed compact subset of state--input space.
We construct a parametric estimator of the map using fixed nonlinear features, so that the model is nonlinear in the state and input, but linear in the unknown parameters. A Gaussian prior over the parameters yields recursive Bayesian posterior updates as data stream in, 
enabling online quantification of predictive uncertainty in the reconstructed dynamics over the target set.
We formulate an optimal adaptive-design problem over an information state, using a prediction-oriented acquisition criterion based on the mean marginal mutual information between candidate future trajectories and the reconstructed dynamics over the target set. 
We then approximate the resulting adaptive-design problem by a non-myopic
receding-horizon formulation, evaluate its remaining expectation using a
scenario-based sample average, and solve the resulting deterministic program
with the cross-entropy method, leveraging parallel candidate--scenario
evaluations.
 Numerical experiments on a noisy multistable system demonstrate that the proposed adaptive information-seeking strategy reduces predictive uncertainty and reconstruction error more efficiently than common excitation baselines under comparable experimental constraints.
\end{abstract}

\section{Introduction}

\begin{figure}[H]
    \centering
\includegraphics[width=\linewidth]{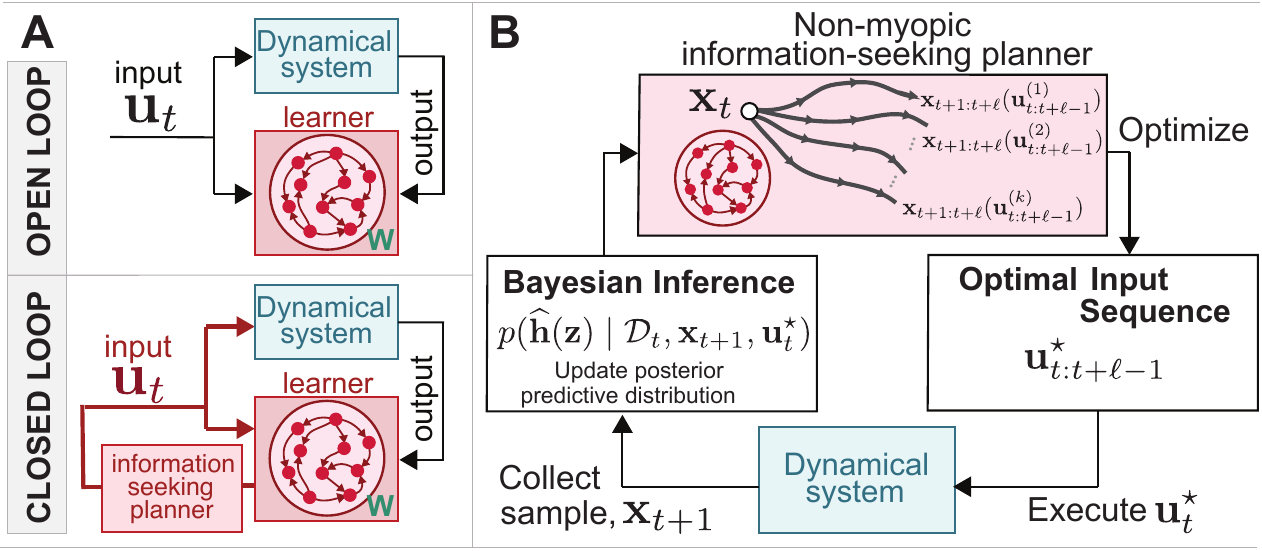}
    \caption{ 
    \small
\textbf{Online system identification and the proposed closed-loop
information-seeking architecture.}
\textbf{(A)} Open-loop (top) and closed-loop (bottom) online system
identification. The learner (red), parameterized by weights $W$,
reconstructs the unknown controlled dynamics from state observations generated by the dynamical system
(blue). In the open-loop setting, the excitation input is specified
independently of the learner. In the closed-loop setting, an
information-seeking planner uses the learner's evolving uncertainty to
adaptively select inputs that produce informative observations.
\textbf{(B)} Receding-horizon Bayesian active-learning loop at time $t$.
Given the current information state
$(\mathbf{x}_t, \mathbf{m}_t,\Omega_t)$, the non-myopic planner evaluates candidate input
sequences and their predicted state trajectories according to their expected
informativeness about the reconstructed dynamics. The optimal input sequence
is selected, but only $\mathbf{u}_t^\star$ is applied to the system.
After observing $\mathbf{x}_{t+1}$, Bayesian inference updates the posterior predictive
distribution, and the cycle is repeated from the resulting updated information
state.}
    \label{fig:schematic}
\end{figure}

Accurate models of nonlinear dynamics are essential for prediction
and control, yet are often unavailable or vary across operating
conditions, individuals, and environments. We therefore consider  the problem of 
\emph{online learning} of stochastic nonlinear dynamical systems
driven by control inputs, where the dynamics are inferred sequentially
from a single observed state trajectory. When the system can be
externally excited, the inputs themselves can be adaptively designed
to acquire informative observations, leading to the problem of
\emph{active learning}---closely related to sequential optimal experimental design \cite{Huan2024} and optimal
exploration \cite{Wagenmaker2023}.  This setting naturally
couples inference and decision-making: model uncertainty is updated
from streaming observations while future inputs are designed to
improve learning. Fig. \ref{fig:schematic}\textbf{A} contrasts the corresponding open- and
closed-loop online identification architectures.

\paragraph*{Related work}
Input design for dynamical system identification has a long history
\cite{Bombois2011,Taylor2021}.
Finite-sample guarantees for learning nonlinear dynamics from trajectory data
have been developed in \cite{Foster2020,Sattar2022}, with active identification
studied in \cite{Mania2022}.
Related approaches include receding-horizon and model-predictive input-design
methods \cite{Hu2026,Toyoda2017,Schultheis2019,Ardekani2025}, greedy
D-optimal exploration (FLEX) \cite{Blanke2023}, GP-based information-seeking
control \cite{Capone2020,Le2021}, optimistic active exploration (OPAX)
\cite{Sukhija2023}, and space-filling input design that promotes coverage of a
prescribed state--input region \cite{KissM.TothR.Schoukens2024}.
A related line instead tailors exploration to downstream control performance; see, e.g., 
\cite{Wagenmaker2023,Lee2024}.
Our active learning approach is \emph{prediction-oriented}: 
we quantify informativeness directly through uncertainty reduction in the reconstructed dynamics
 over a prescribed region of state-input space, connecting to
goal-oriented Bayesian experimental design \cite{Zhong2026} and expected
predictive information gain \cite{Smith2023}.
Policy-based Bayesian experimental-design methods such as DAD
\cite{Foster2021DAD} approximate the adaptive design policy directly; in
contrast, we solve a constrained non-myopic receding-horizon stochastic input-design problem online and replan with an updated model after each observation.

\paragraph*{Main contributions}
We introduce a prediction-oriented Bayesian active-learning framework for the
online identification of nonlinear stochastic dynamical systems over a
prescribed region of state--input space. First, we formulate informativeness
directly in terms of the reconstructed dynamics, using a mutual
information criterion that quantifies the expected reduction in predictive
uncertainty over the target region.
Second, we cast adaptive input design as a sequential decision problem over an information state comprising the physical state and the current Bayesian
posterior, and introduce a constrained, non-myopic receding-horizon approximation to the resulting causal design problem (see Fig. \ref{fig:schematic}\textbf{B}).
Third, for a nonlinear feature model that is linear in the unknown
parameters, the Bayesian updates are recursive and the predictive information
gain is available analytically; the remaining expectation over uncertain
future trajectories is approximated by a scenario-based sample average and
optimized using CEM. We illustrate the resulting closed-loop
information-seeking strategy on a noisy multistable system, where it reduces
predictive uncertainty and reconstruction error more efficiently than
power-matched excitation baselines.

\paragraph*{Paper structure}
The remainder of the paper is organized as follows.  \S \ref{sec:mathematical_prel} introduces 
mathematical preliminaries.  \S\ref{sec:model_dynamics} presents the model of the controlled
dynamics and the learning objective, and \S\ref{sec:Bayesian_learning} develops the corresponding
Bayesian inference framework. \S\ref{sec:planning} formulates the prediction-oriented
information objective and the receding-horizon stochastic input-design problem,
together with its scenario-based approximation. \S\ref{sec:algorithms} summarizes the
resulting algorithm, \S\ref{sec:numerical_case_study} presents a numerical case study, and
\S\ref{sec:conclusion} concludes.

\section{Mathematical Preliminaries}
\label{sec:mathematical_prel}

\paragraph{Notation}
We use $a \in \RR$ for scalars, $\va\in \RR^k$ for vectors and $A\in \RR^{n \times m}$ for matrices. 
$A\trp$ denotes the transpose of $A$. $\mathbb{I}_n$ denotes the $n$-dimensional identity matrix. The symbol $:=$ denotes equality by definition. We use $\va_{t:t+\ell}:= (\va_t, \va_{t+1}, \dots, \va_{t+\ell})$ for temporal sequences.  The superscript $^\star$ denotes optimality---not to be confused with $^*$, which we use to denote quantities evaluated over a set of grid points. Given a matrix $A\in \RR^{n \times m}$, $\va := \text{vec}(A) \in \RR^{nm}$ denotes the vectorization of $A$, obtained by stacking the columns of $A$ on top of one another.
A useful identity is $\text{vec}(A\vx) = (\mathbb{I}_n \otimes \vx\trp) \text{vec}(A\trp) = (\vx\trp \otimes \mathbb{I}_n) \text{vec}(A)$, where $\otimes$ is the Kronecker product.

\paragraph{Definitions}  
Let $\vx$ be a continuous random vector supported on $\mX\subseteq \RR^n$ with density $p_{\vx}$. Its \textit{differential entropy} is defined as
$
    \mathcal{H}(\vx)
    :=
    - \int_{\mX} p_{\vx}(\vz) \ln p_{\vx}(\vz) \, d\vz
    =
    - \E\!\left[\ln p_{\vx}(\vx)\right].
    \label{def:diff_entropy}
$
The differential entropy quantifies the average uncertainty associated with $\vx$.
If
$\vx \sim \Nrm(\vmu,\Sigma)$ with 
$\Sigma \succ 0$, then
$
    \mathcal{H}(\vx)
    =
    \frac{1}{2}\ln\!\big((2\pi e)^n \det(\Sigma)\big).
$
The \textit{conditional differential entropy} of a random vector $\vy$ given a random vector $\vx$ is defined by
$
   \mathcal{H}(\vy \,|\, \vx) := - \int_\mX \int_\mathcal{Y} p(\vy,\vx) \ln p(\vy\,|\,\vx) d \vy d\vx.
\label{eq:conditional_entropy}
$

Finally, the \textit{mutual information between the random vectors} $\vx$ and $\vy$ 
can be written as
\begin{equation*}
    \mathsf{MI}(\vx; \vy) = \mathcal{H}(\vx) - \mathcal{H}(\vx \,|\, \vy) = \mathcal{H}(\vy) - \mathcal{H}(\vy \,|\, \vx).
    \label{eq:MI_def_2}
\end{equation*}
From a Bayesian perspective, 
we can view $p(\vx)$ as the \textit{prior} distribution for $\vx$ and $p(\vx\,|\,\vy)$ as the \textit{posterior} distribution after observation of data $\vy$. 
The $\text{MI}(\vx; \vy)$ therefore  measures the expected reduction in the entropy of \(\vx\) upon observing $\vy$. This interpretation is central to the present work.

\section{Model of the System Dynamics}
\label{sec:model_dynamics}

We consider the online identification of a discrete-time stochastic
controlled Markov process of the form
\begin{equation}
    \mathbf{x}_{t+1}
    =
    \mathbf{x}_t
    +
    \mathbf{h}(\mathbf{x}_t,\mathbf{u}_t)
    +
    \boldsymbol{\xi}_t,
    \label{eq:dynamics}
\end{equation}
 where $\mathbf{x}_t\in\mathbb{R}^{n}$ is the system state,
$\mathbf{u}_t\in\mathbb{R}^{m}$ is the control input, and
$
    \boldsymbol{\xi}_t
    \overset{\mathrm{i.i.d.}}{\sim}
    \mathcal{N}(\mathbf{0},\Sigma_{\xi})
    \text{ with }
    \Sigma_{\xi}\succ0,
    \label{eq:process_noise}
$
assumed known.
The \textit{unknown} map
$\mathbf{h}:\mathbb{R}^{n+m}\rightarrow\mathbb{R}^{n}$
denotes the deterministic one-step state increment induced jointly
by the current state and control input. Defining the augmented
state--input variable
$
    \mathbf{z}_t
    :=
    \begin{bmatrix}
        \mathbf{x}_t^\top &
        \mathbf{u}_t^\top
    \end{bmatrix}^{\top}
    \in\mathbb{R}^{n+m},
    \label{eq:z_definition}
$
we use the shorthand
$\mathbf{h}(\mathbf{z}_t)
:=\mathbf{h}(\mathbf{x}_t,\mathbf{u}_t)$. Our objective is to learn the controlled state-increment map over a
prescribed compact region of interest (ROI)
$\mathcal{Z}\subset\mathbb{R}^{n+m}$ from a single controlled
trajectory of finite length $T$. Thus, the learning target is the
restriction $\mathbf{h}|_{\mathcal{Z}}$, rather than the parameters
of a particular representation of $\mathbf{h}$.

We model the unknown dynamics using a \textit{fixed nonlinear feature map}
$
    \boldsymbol{\phi}:
    \mathbb{R}^{n+m}\rightarrow\mathbb{R}^{d_\phi}
$
and a matrix of unknown weights
$W\in\mathbb{R}^{d_\phi\times n}$, defining the parametric estimator
$
    \widehat{\mathbf{h}}(\mathbf{z};W)
    :=
    W^\top\boldsymbol{\phi}(\mathbf{z}).
    \label{eq:feature_model}
$
This class of models can capture many types of dynamics and is used widely in system identification \cite{Mania2022,Wagenmaker2023}. 
Let $\mathbf{w}:=\operatorname{vec}(W)\in\mathbb{R}^{nd_\phi}$ and define
$
    \Phi(\mathbf{z})
    := \mathbb{I}_n\otimes\boldsymbol{\phi}(\mathbf{z})^\top
    \in\mathbb{R}^{n\times nd_\phi}.
    \label{eq:Phi_definition}
$
Then
$
    \widehat{\mathbf{h}}(\mathbf{z};\mathbf{w})
    =
    \Phi(\mathbf{z})\mathbf{w},
    \label{eq:parametric_h}
$
and the corresponding reconstructed dynamics model is
\begin{equation}
    \mathbf{x}_{t+1}
    =
    \mathbf{x}_t
    +
    \Phi(\mathbf{x}_t, \mathbf{u}_t )\mathbf{w}
    +
    \boldsymbol{\xi}_t.
    \label{eq:parametric_dynamics}
\end{equation}
\begin{ass}[Full-state measurements]
    \label{ass:measurements}
    The state $\mathbf{x}_t$ is observed without measurement noise at
each sampling instant $t$.
\end{ass}

\begin{ass}[Realizability]
\label{ass:realizability}
There exists
$\mathbf{w}^{\dagger}\in\mathbb{R}^{nd_\phi}$ such that
$
    \mathbf{h}(\mathbf{z})
    =
    \Phi(\mathbf{z})\mathbf{w}^{\dagger}
$
for all state--input pairs $\mathbf{z}$ considered during learning and planning, including
all $\mathbf{z}\in\mathcal{Z}$.
\end{ass}

\paragraph*{Feature $\vphi$ choice} 
In practice, we use a random-feature model with single-hidden-layer features of the form
$
    \boldsymbol{\phi}(\mathbf{z})
    =
    \tanh(J\mathbf{z}+\mathbf{b}),
    \label{eq:random_features}
$
where
$J\in\mathbb{R}^{d_\phi\times(n+m)}$ and
$\mathbf{b}\in\mathbb{R}^{d_\phi}$ are sampled before learning
starts and subsequently held fixed; only the output weights $W$ are
inferred. This corresponds to a random-hidden-layer
(ELM/RVFL) model \cite{Huang2006,Igelnik1995}. 
Finite linear combinations of sigmoidal features are dense in spaces of continuous functions on compact domains \cite{Cybenko1989,Leshno1993}, while related results establish universal approximation for random-hidden-layer models under suitable sampling schemes \cite{Igelnik1995}.
The feature distribution should be scaled to the characteristic
length scales of $\mathcal{Z}$ so as to avoid widespread saturation;
the construction used in our experiments is described in
\hyperref[subsec:feature_selec_recipe]{Appendix A}.

\paragraph*{Learning objective}
Given a finite learning horizon $T$, our goal is to accurately reconstruct
the restriction $\mathbf{h}\vert_{\mathcal{Z}}$ from the sequentially collected data.
To this end, we maintain a posterior distribution over the weights
$\mathbf{w}$, which induces a posterior predictive distribution over
$\widehat{\mathbf{h}}|_{\mathcal{Z}}$. 
This domain-level objective distinguishes dynamical-system
reconstruction from merely fitting the observed trajectory \cite{Hess2024}.
When a finite-dimensional
representation of the ROI is required computationally, we denote by
$
    \mathcal{Z}^{\ast}
    :=
    \big\{
        \mathbf{z}^{(i)}_{\ast}
    \big\}_{i=1}^{N_{\ast}}
    \subset\mathcal{Z}
$
the corresponding evaluation grid with a total of $N_*$ grid points.

The Bayesian posterior predictive distribution underlying the learning and planning components of the active learning method is developed next.

\section{Bayesian learning }
\label{sec:Bayesian_learning}

At time $t$, given the dataset
$\mathcal{D}_t:=\{\mathbf{x}_{0:t},\mathbf{u}_{0:t-1}\}$,
we maintain a Gaussian posterior over the unknown weights,
$
    p(\mathbf{w}\, | \, \mathcal{D}_t)
    =
    \mathcal{N}(\mathbf{m}_t, \Omega_t).
$
The linear-in-parameters dynamics and Gaussian additive process noise yield
\textit{closed-form Bayesian linear-regression updates}: recursive updates are
used to assimilate streaming observations, while their batch
counterpart is used to evaluate hypothetical multi-step updates during
planning. For fixed $d_\phi$, all previously collected data are
summarized by the fixed-dimensional sufficient statistics
$(\mathbf{m}_t, \Omega_t)$.

\paragraph{Prior over $\mathbf{w}$}
At $t=0$, we place the Gaussian prior
$
    \mathbf{w}
    \sim
    \mathcal{N}(\mathbf{m}_0, \Omega_0),
$
with
$
    \Omega_0 \succ 0.
$
$(\mathbf{m}_0, \Omega_0)$ encode prior knowledge
and regularize estimation in the small-data regime.

\paragraph{Posterior updates over $\mathbf{w}$}
Define the observed state increment
$
    \mathbf{y}_{t+1}
    :=
    \mathbf{x}_{t+1}-\mathbf{x}_t,
$
so that, from~\eqref{eq:parametric_dynamics},
$
    \mathbf{y}_{t+1}
    =
    \Phi(\mathbf{x}_t,\mathbf{u}_t)\mathbf{w}
    +
    \boldsymbol{\xi}_t.
$
For notational compactness, let
$\Phi_{\tau-1}:=
\Phi(\mathbf{x}_{\tau-1},\mathbf{u}_{\tau-1})$.
Starting from
$p(\mathbf{w}\mid\mathcal{D}_t)
=\mathcal{N}(\mathbf{m}_t, \Omega_t)$
and assimilating a batch of \textit{$\ell$ additional observations} yields
\begin{subequations}
\label{eq:batch_update}
\begin{align}
    \Omega_{t+\ell\mid t}^{-1}
    &=
    \Omega_t^{-1}
    +
    \sum_{\tau=t+1}^{t+\ell}
    \Phi_{\tau-1}^{\top}
    \Sigma_{\xi}^{-1}
    \Phi_{\tau-1},
    \label{eq:batch_cov}\\
    \mathbf{m}_{t+\ell\mid t}
    &=
    \Omega_{t+\ell\mid t}
    \left(
        \Omega_t^{-1}\mathbf{m}_t
        +
        \sum_{\tau=t+1}^{t+\ell}
        \Phi_{\tau-1}^{\top}
        \Sigma_{\xi}^{-1}
        \mathbf{y}_{\tau}
    \right).
    \label{eq:batch_mean}
\end{align}
\end{subequations}
The derivation is given in 
\hyperref[subsec:app_posterior_over_w]{Appendix B}. 

For \textit{online assimilation ($\ell = 1$)}, after observing
$(\mathbf{x}_t,\mathbf{u}_t,\mathbf{x}_{t+1})$, define the Kalman gain
$
    K_{t+1}
    :=
    \Omega_t\Phi_t^\top
    \left(
        \Sigma_{\xi}
        +
        \Phi_t \Omega_t\Phi_t^\top
    \right)^{-1}.
$
The recursive posterior update is 
\begin{subequations}
\label{eq:recursive_update}
\begin{align}
    \mathbf{m}_{t+1}
    &=
    \mathbf{m}_t
    +
    K_{t+1}
    \left(
        \mathbf{y}_{t+1}-\Phi_t\mathbf{m}_t
    \right),
    \label{eq:recursive_mean}\\
    \Omega_{t+1}
    &=
    \Omega_t
    -
    K_{t+1}\Phi_t \Omega_t .
    \label{eq:recursive_cov}
\end{align}
\end{subequations}

\paragraph{Posterior predictive distribution}
The posterior over $\mathbf{w}$ induces, at every
$\mathbf{z}\in\mathcal{Z}$, the Gaussian posterior predictive distribution
\begin{equation}
    p\Big(
        \widehat{\mathbf{h}}(\mathbf{z})
        \mid\mathcal{D}_t \Big)
    =
    \mathcal{N}
    \left(
        \Phi(\mathbf{z})\mathbf{m}_t,\,
        C_t(\mathbf{z})
    \right),
    \label{eq:predictive}
\end{equation}
where
$
    C_t(\mathbf{z})
    :=
    \Phi(\mathbf{z})
    \Omega_t
    \Phi(\mathbf{z})^\top
$
quantifies the local epistemic uncertainty in the reconstructed
controlled state-increment map.

Importantly, our learning objective is \textit{goal-oriented} and, more
specifically, \textit{prediction-oriented}: $\mathbf{w}$ is an intermediate
representation, whereas the quantity of interest is
$\widehat{\mathbf{h}}|_{\mathcal{Z}}$.
Reducing uncertainty in $\mathbf{w}$ need not imply a commensurate
reduction in uncertainty in the reconstructed dynamics over the target
region, since different parameter directions can have different
predictive relevance on $\mathcal{Z}$.
This perspective connects to goal-oriented Bayesian experimental
design~\cite{Zhong2026} and, in particular, to prediction-oriented
Bayesian active learning and its \textit{expected predictive information gain}
(EPIG) criterion \cite{Smith2023}, which targets information
about predictions rather than model parameters.
In \S\ref{sec:planning}, we develop a multi-step dynamical analogue in which
candidate control sequences are evaluated by the information their
induced future trajectories are expected to provide about
$\widehat{\mathbf{h}}(\mathbf{z})$ across $\mathcal{Z}$.

\section{Information-seeking planning}
\label{sec:planning}

We now turn to the \textit{planning} component of the active-learning loop in
Fig. \ref{fig:schematic}\textbf{B}. The objective is to design control inputs that steer the
system toward observations that are informative about the controlled
state-increment map over the target region $\mathcal{Z}$. We first
define the prediction-oriented information criterion to be used as acquisition function, then formulate
the adaptive design problem over causal feedback policies, and
finally introduce the receding-horizon and sample-average
approximations used in our implementation.

\subsection{Prediction-oriented information objective}
\label{subsec:acquisition}

As discussed in \S\ref{sec:Bayesian_learning}, the quantity of interest is the restriction of the reconstructed map $\widehat{\mathbf{h}}|_{\mathcal{Z}}$. At a target location
$\mathbf{z}\in\mathcal{Z}$ and time $t$, its uncertainty is quantified by
the differential entropy of its posterior predictive distribution 
\begin{equation*}
    \mathcal{H} \!\left(
        \widehat{\mathbf h}(\mathbf z)\mid\mathcal D_t
    \right)
    =
    \frac{1}{2}
    \ln\!\left[
        (2\pi e)^n
        \det C_t(\mathbf z)
    \right],
    \label{eq:local_entropy}
\end{equation*}
where $C_t(\mathbf z)$ is given in~\eqref{eq:predictive}.

Let $d$ denote a prospective 
design and let $\mathcal Y_d$ denote the future state observations
generated under that design. 
Let $\rho$ be a probability density over
$\mathcal Z$, describing the relative importance assigned to different
parts of the learning region. We define the \textit{mean marginal mutual
information} ($\mathsf{mmMI}$) of the design as \cite[\S 4.1]{MacKay1992} 
\begin{equation}
    \mathsf{mmMI}_t(d)
    :=
    \int_{\mathcal Z}
    \rho(\mathbf z)\,
    \mathsf{MI}\!\left(
        \mathcal Y_d;
        \widehat{\mathbf h}(\mathbf z)
        \mid
        \mathcal D_t,d
    \right)
    d\mathbf z.
    \label{eq:mmmi_continuous}
\end{equation}
That is, $\mathsf{mmMI}_t(d)$ is the expected reduction in
posterior predictive entropy induced by the candidate experiment,
averaged over the region $\mathcal{Z}$ where accurate reconstruction matters. 

For numerical evaluation, we discretize the integral over
$\mathcal Z$ using the grid
$\mathcal Z^\ast
=\{\mathbf z_\ast^{(i)}\}_{i=1}^{N_\ast}$
introduced in \S\ref{sec:model_dynamics}. Let $\omega_i\geq0$,
$\sum_{i=1}^{N_\ast}\omega_i=1$, denote the associated quadrature
weights. We define the grid approximation
\begin{equation}
    \mathsf{mmMI}^{\ast}_t(d)
    :=
    \sum_{i=1}^{N_\ast}
    \omega_i\,
    \mathsf{MI}\!\left(
        \mathcal Y_d;
        \widehat{\mathbf h}(\mathbf z_\ast^{(i)})
        \mid
        \mathcal D_t,d
    \right).
    \label{eq:mmmi_grid}
\end{equation}

For the linear-in-parameters Gaussian model of \S\ref{sec:model_dynamics}-\ref{sec:Bayesian_learning}, the \textit{marginal mutual
information at each target location}, $ \mathsf{mMI}_t(d,\mathbf z)
    :=
    \mathsf{MI}\!\left(
        \mathcal Y_d;
        \widehat{\mathbf h}(\mathbf z)
        \mid
        \mathcal D_t,d
    \right),$ admits the following analytic log-determinant representation
\begin{equation}
    \mathsf{mMI}_t(d,\mathbf z)
    =
    \frac{1}{2}
    \mathbb E
    \left[
        \ln
        \frac{
            \det C_t(\mathbf z)
        }{
            \det C_{t\mid\mathcal Y_d}(\mathbf z)
        }
    \right],
    \label{eq:local_mmi_logdet}
\end{equation}
where $C_{t\mid\mathcal Y_d}(\mathbf z)$ denotes the predictive
covariance after assimilating the hypothetical observations
$\mathcal Y_d$, and the expectation is taken with respect to their
predictive distribution under design $d$.

\subsection{Optimal adaptive design \& receding-horizon planning}
\label{subsec:causal_to_rh}

Let
$
    b_t(\mathbf{w})
    :=
    p(\mathbf{w}\, | \, \mathcal{D}_t)
$
denote the posterior belief over the weights $\mathbf{w}$ at time $t$. Under the Gaussian model of
\S\ref{sec:Bayesian_learning}, $b_t$ is completely specified by
$(\mathbf{m}_t, \Omega_t)$. Hence, 
\begin{equation*}
    \mathbf{s}_t
    :=
    (\mathbf{x}_t,b_t)
    \equiv
    (\mathbf{x}_t,\mathbf{m}_t,\Omega_t)
    \label{eq:information_state}
\end{equation*}
forms a sufficient \textit{information state} for sequential decision-making:
the physical state specifies the current system configuration, while
the posterior summarizes the information contained in all previous
observations 
\cite{Kappen2011,ShenHuan2023,kumar2016stochastic}.
Accordingly, under the physical dynamics \eqref{eq:dynamics} and Bayesian update \eqref{eq:recursive_update}, the information-state process is controlled Markov.

Let $\mathcal U\subset\mathbb R^m$ denote the admissible input set.
A \textit{causal design policy}
$\boldsymbol{\pi}_{t:T-1}:=\{\pi_t,\ldots,\pi_{T-1}\}$ consists of decision rules
$\pi_\tau:\mathcal S\rightarrow\mathcal U$ such that
\begin{equation*}
    \mathbf u_\tau=\pi_\tau(\mathbf s_\tau),
    \qquad \tau=t,\ldots,T-1,
\end{equation*}
where $\mathcal S$ denotes the information state-space. Thus, each
input may depend on information available up to the current time, but
not on future observations.

Specializing the prediction-oriented criterion \eqref{eq:mmmi_continuous} to the causal design policy 
$d=\boldsymbol{\pi}_{t:T-1}$ and conditioning on the information-state $\mathbf{s}_t$ at time $t$, the corresponding
remaining-horizon objective is
\begin{equation*}
   J_t(\mathbf{s}_t; \boldsymbol{\pi}_{t:T-1}) := \int_{\mathcal{Z}}
\rho(\mathbf{z})\, 
\mathsf{MI}\!\left(
    \widehat{\mathbf{h}}(\mathbf{z});
    \mathbf{x}_{t+1:T}
    \,|\,
    \mathbf{s}_t,\boldsymbol{\pi}_{t:T-1}
\right)
d\mathbf{z}.
\end{equation*}
Accordingly, the remaining-horizon active-design problem is
\begin{equation}
\tag{OPT1}\label{opt:causal_policy}
\begin{aligned}
\boldsymbol{\pi}_{t:T-1}^{\star}
\in
\arg\max_{\boldsymbol{\pi}_{t:T-1}}
 &
 J_t(\mathbf{s}_t; \boldsymbol{\pi}_{t:T-1})
\\
\mathrm{s.t.}\quad&
\mathbf{u}_{\tau}
=
\pi_{\tau}(\mathbf{s}_{\tau})
\in\mathcal{U},
\;\;
\tau=t,\ldots,T-1,
\end{aligned}
\end{equation}
where the information state evolves according to the physical
dynamics~\eqref{eq:dynamics} and the Bayesian update~\eqref{eq:recursive_update}. Thus, \eqref{opt:causal_policy} seeks a causal feedback
policy that maximizes the information acquired about the reconstructed
dynamics over the remainder of the experiment.

By the chain rule for mutual information and sufficiency of the information state, the objective in \eqref{opt:causal_policy}
admits an additive representation in terms of successive conditional mutual-information terms. Define the one-step mutual-information reward
$
r(\mathbf{s}, \mathbf{u}):= \int_{\mathcal{Z}} \rho(\mathbf{z}) \, \mathsf{MI}\big(\widehat{\mathbf{h}}(\mathbf{z}); \mathbf{x}^+ \, | \, \mathbf{s}, \mathbf{u}\big) d\mathbf{z},
$
where $\mathbf{x}^+$ denotes the next state. Then, for any causal policy $\boldsymbol{\pi}_{t:T-1}$,
$
J_t(\mathbf{s}_t;\boldsymbol{\pi}_{t:T-1}) =  \mathbb{E}^{\boldsymbol{\pi}}\big[ \sum_{\tau=t}^{T-1} r\big(\mathbf{s}_\tau, \boldsymbol{\pi}_\tau(\mathbf{s}_\tau)\big) \mid \mathbf{s}_t \big],
$ where $\mathbb{E}^{\boldsymbol{\pi}}$ denotes expectation under policy $\boldsymbol{\pi}_{t:T-1}$.
 Together with the controlled-Markov property of
$\mathbf{s}_\tau$, this allows the exact adaptive-design problem to be formulated
through a Bellman recursion over the information state:
$
V_t(\mathbf{s}) = \max_{\mathbf{u} \in \mathcal{U}} \, \big\{ r(\mathbf{s}, \mathbf{u}) + \mathbb{E}[V_{t+1}(\mathbf{s}^+) \, \vert \, \mathbf{s}, \mathbf{u}]\big\},
$
where $\mathbf{s}^+$ denotes the next information state and $V_t(\mathbf{s})$ is 
the optimal expected cumulative information reward from time $t$ to $T$, with $V_T(\mathbf{s}) = 0$.
Exact dynamic
programming is, however, generally impractical here because the
information state $(\mathbf{x}_\tau, \mathbf{m}_\tau,\Omega_\tau)$ is continuous and
high-dimensional, the system dynamics are nonlinear and stochastic,
and the inputs are constrained. Related policy-based Bayesian
experimental-design methods parameterize and approximate the
adaptive design policy directly \cite{Foster2021DAD,Hedman2025StepDAD}.
Instead, we use a \textit{receding-horizon} approximation. At each realized information state
$\mathbf{s}_t$, we replace the remaining causal policy by a finite
open-loop control sequence of length $\ell$,
\begin{equation*}
    \mathbf{U}^{(\ell)}_t
    :=
    \left(
        \mathbf{u}_{t|t},
        \ldots,
        \mathbf{u}_{t+\ell-1|t}
    \right),
    \qquad 1 \leq \ell\leq T-t.
    \label{eq:planned_input_sequence}
\end{equation*}

Specializing the $\mathsf{mmMI}$ criterion \eqref{eq:mmmi_continuous}  to the finite open-loop design \(d=\mathbf U_t^{(\ell)}\), define
\begin{equation*}
\begin{split}
    J_t^{(\ell)}
    \!\left(
        \mathbf{U}^{(\ell)}_t;
        \mathbf{s}_t
    \right)
    :=
    \int_{\mathcal{Z}}
    \rho(\mathbf{z})\,
    \mathsf{MI}\!\Big(
        \widehat{\mathbf{h}}(\mathbf{z});
        \mathbf{x}_{t+1:t+\ell}
        \,\Big|\,
        \mathbf{s}_t,
        \mathbf{U}^{(\ell)}_t
    \Big)
    d\mathbf{z}.
    \label{eq:rh_mmmi}
\end{split}
\end{equation*}
Accordingly, the \textit{stochastic receding-horizon design problem} is
\begin{equation}
\tag{OPT2}\label{opt:stochastic_mpc}
\begin{aligned}
\mathbf{U}^{(\ell)\star}_t
\in
\arg\max_{\mathbf{U}^{(\ell)}_t}
\quad&
J_t^{(\ell)}
\!\left(
    \mathbf{U}^{(\ell)}_t;
    \mathbf{s}_t
\right)
\\
\mathrm{s.t.}\quad&
\mathbf{u}_{\tau|t}\in\mathcal{U},
\qquad
\tau=t,\ldots,t+\ell-1,
\end{aligned}
\end{equation}
where the distribution of the hypothetical trajectories entering \(J_t^{(\ell)}\) is induced by the dynamics \eqref{eq:dynamics} and Bayesian updates \eqref{eq:recursive_update}, initialized at the current information state $\mathbf{s}_t$.

At each $t$, only the first optimized input is applied,
$\mathbf{u}_t=\mathbf{u}^{\star}_{t|t}$. After observing
$\mathbf{x}_{t+1}$, the posterior is updated and~\eqref{opt:stochastic_mpc}
is solved again from the newly realized information state. Thus, each individual planning problem is open loop over its prediction horizon, while
repeated replanning induces an implicit closed-loop policy. 
The non-myopic planning horizon $\ell>1$ permits the planner to
steer toward informative regions that may not be reached greedily,
while limiting the online computational cost.

\subsection{Sample-average approximation}
\label{subsec:saa}

Problem~\eqref{opt:stochastic_mpc} is a nonlinear stochastic
program.
We approximate its
remaining expectation by a scenario-based sample-average approximation
(SAA). 

At each replanning time $t$, draw
$
    \mathbf{w}^{(j)}
    \sim
    \mathcal{N}(\mathbf{m}_t, \Omega_t),
    \;
    j=1,\ldots,N_w,
$
and independently draw $N_{\xi}$ process-noise sequences
$
    \boldsymbol{\xi}_{t:t+\ell-1}^{(k)}
    :=
    \left(
        \boldsymbol{\xi}_{t|t}^{(k)},
        \ldots,
        \boldsymbol{\xi}_{t+\ell-1|t}^{(k)}
    \right)
    \text{ with }
    \boldsymbol{\xi}_{\tau|t}^{(k)}
    \overset{\mathrm{i.i.d.}}{\sim}
    \mathcal{N}(\mathbf{0}, \Sigma_{\xi}),
$
for $k=1,\ldots,N_{\xi}$. We use the Cartesian product of these
samples, yielding $N_wN_{\xi}$ rollout scenarios. For each pair
$(j,k)$ and candidate sequence $\mathbf{U}^{(\ell)}_t$, set
$\mathbf{x}_{t|t}^{(j,k)}=\mathbf{x}_t$ and propagate
\begin{equation*}
    \mathbf{x}_{\tau+1|t}^{(j,k)}
    =
    \mathbf{x}_{\tau|t}^{(j,k)}
    +
    \Phi\!\left(
        \mathbf{x}_{\tau|t}^{(j,k)},
        \mathbf{u}_{\tau|t}
    \right)
    \mathbf{w}^{(j)}
    +
    \boldsymbol{\xi}_{\tau|t}^{(k)},
    \label{eq:saa_rollout}
\end{equation*}
for $\tau=t,\ldots,t+\ell-1$. Denote
$
    \Phi_{\tau|t}^{(j,k)}
    :=
    \Phi\!\left(
        \mathbf{x}_{\tau|t}^{(j,k)},
        \mathbf{u}_{\tau|t}
    \right).
$
Then, the corresponding hypothetical terminal posterior covariance satisfies
\begin{equation*}
\left(
    \Omega_{t+\ell|t}^{(j,k)}
\right)^{-1}
=
\Omega_t^{-1}
+
\sum_{\tau=t}^{t+\ell-1}
\Phi_{\tau|t}^{(j,k)\top}
\Sigma_{\xi}^{-1}
\Phi_{\tau|t}^{(j,k)}.
\label{eq:saa_covariance}
\end{equation*}

Conditioned on the current information state $\mathbf{s}_t=(\mathbf{x}_t,\mathbf{m}_t, \Omega_t)$,
using the empirical average over the sampled trajectories to approximate the acquisition function yields the
\textit{deterministic SAA problem}
\begin{equation}
\tag{OPT3}\label{opt:saa_mpc}
\begin{aligned}
\min_{\mathbf{U}^{(\ell)}_t} \;\;  & 
\frac{1}{N_wN_{\xi}}
\sum_{j=1}^{N_w}
\sum_{k=1}^{N_{\xi}}
\sum_{i=1}^{N_{\ast}}
\omega_i
\ln\det\!\left(
    \Phi_i^{\ast}
    \Omega_{t+\ell|t}^{(j,k)}
    \Phi_i^{\ast\top}
\right) 
\\
\mathrm{s.t.}\quad&
\mathbf{x}_{t|t}^{(j,k)}
=
\mathbf{x}_t,
\\
&
\mathbf{x}_{\tau+1|t}^{(j,k)}
=
\mathbf{x}_{\tau|t}^{(j,k)}
+
\Phi_{\tau|t}^{(j,k)}
\mathbf{w}^{(j)}
+
\boldsymbol{\xi}_{\tau|t}^{(k)},
\\
&
\left(
    \Omega_{t+\ell|t}^{(j,k)}
\right)^{-1}
=
\Omega_t^{-1}
+
\sum_{\tau=t}^{t+\ell-1}
\Phi_{\tau|t}^{(j,k)\top}
\Sigma_{\xi}^{-1}
\Phi_{\tau|t}^{(j,k)},
\\
&
\mathbf{u}_{\tau|t}\in\mathcal{U},
\qquad
\tau=t,\ldots,t+\ell-1,
\\
&
j=1,\ldots,N_w,\qquad
k=1,\ldots,N_{\xi},
\end{aligned}
\end{equation}
where 
$
    \Phi_i^\ast := \Phi(\mathbf z_\ast^{(i)}), \, i=1,\ldots,N_\ast .
$
We denote the objective in \eqref{opt:saa_mpc} by $\mathcal{C}_t^{(\ell)}\big( \mathbf{U}_t^{(\ell)}; \mathbf{s}_t\big)$.

Finally, we note that a generic expected-information-gain objective may  require a
nested Monte Carlo calculation to estimate the information criterion,
which can lead to unfavorable statistical and computational properties
\cite[\S3]{Rainforth2024a}. Here, by contrast,
the linear-in-parameters Gaussian structure provides the predictive information gain
analytically through the log-determinant expression
\eqref{eq:local_mmi_logdet}; Monte Carlo is used only for the outer
expectation over uncertain future trajectories.
In \S\ref{sec:algorithms} we solve problem \eqref{opt:saa_mpc} online using
the cross-entropy method (CEM).

\section{Algorithm for receding-horizon Bayesian active learning}
\label{sec:algorithms}

The complete receding-horizon Bayesian active-learning 
framework is summarized in Algorithm \ref{alg:rh_active_learning}, with the CEM solver for
the deterministic SAA problem \ref{opt:saa_mpc} detailed in
Algorithm \ref{alg:cem_mpc_scenario}. At each replanning time, a common batch of
$N_wN_\xi$ parameter--noise scenarios is sampled once and reused
to score all $P$ candidate input sequences throughout the CEM iterations.
The candidate--scenario rollouts can be
evaluated in parallel (across CPU threads or GPU cores). CEM iteratively samples input sequences,
retains the lowest-cost elite set, and refits its Gaussian sampling
distribution. After each planning solve, the final sampling distribution is
shifted one step forward to warm-start the next solve. Only the first
input of the best sequence is applied before the posterior is updated
and planning is repeated---see Fig. \ref{fig:schematic}\textbf{B}.

\begin{algorithm}
\caption{Receding-horizon active learning}
\small
\label{alg:rh_active_learning}
\begin{algorithmic}[1]
\REQUIRE Learning horizon $T$; maximum planning horizon $\ell_m$;
feature map $\Phi$; 
initial information state $\mathbf{s}_0=(\mathbf{x}_0, \mathbf{m}_0, \Omega_0)$; CEM parameters $\Theta_{\mathrm{CEM}}$

\STATE Initialize CEM distribution $(\boldsymbol{\mu}_U, \boldsymbol{\sigma}_U)$

\FOR{$t=0,\ldots,T-1$}

    \STATE $\ell\gets\min\{\ell_m,T-t\}$

    \STATE $[\mathbf{U}_t^{(\ell)\star},\boldsymbol{\mu}_U, \boldsymbol{\sigma}_U]
    \gets
    \textsc{CEM-Plan}
    (\mathbf{s}_t,
    \ell,\boldsymbol{\mu}_U, \boldsymbol{\sigma}_U;\Theta_{\mathrm{CEM}})$

    \STATE $\mathbf{u}_t\gets\mathbf{u}_{t|t}^{\star}$

    \STATE Apply $\mathbf{u}_t$ and observe $\mathbf{x}_{t+1}$

    \STATE $\mathbf{y}_{t+1}
    \gets
    \mathbf{x}_{t+1}-\mathbf{x}_t$

    \STATE Update
    $(\mathbf{m}_{t+1},\Omega_{t+1})$
    using~\eqref{eq:recursive_update}
\ENDFOR
\RETURN $\mathbf{m}_T, \Omega_T$
\end{algorithmic}
\end{algorithm}

\begin{algorithm}
\caption{CEM solver for SAA planning (CEM-PLAN)}
\label{alg:cem_mpc_scenario}
\small
\begin{algorithmic}[1]

\REQUIRE Current information state 
$(\mathbf{x}_t, \mathbf{m}_t,\Omega_t)$; horizon $\ell$;
CEM distribution $(\boldsymbol{\mu}_U,\boldsymbol{\sigma}_U)$;
$P$ candidates; $P_e$ elites; $N_{\mathrm{CEM}}$ iterations;
$N_w,N_\xi$; minimum standard deviation $\sigma_{\min}$;
smoothing parameter $\beta$

\STATE  $(\boldsymbol{\mu}_U,\boldsymbol{\sigma}_U)
\gets
(\boldsymbol{\mu}_U(:,1:\ell),\boldsymbol{\sigma}_U(:,1:\ell))$

\STATE Draw
$\mathbf{w}^{(j)}
\sim\mathcal{N}(\mathbf{m}_t, \Omega_t)$,
$j=1,\ldots,N_w$

\STATE Draw
$\boldsymbol{\xi}^{(k)}_{t:t+\ell-1}$,
$k=1,\ldots,N_\xi$, with
$\boldsymbol{\xi}^{(k)}_{\tau|t}
\overset{\mathrm{i.i.d.}}{\sim}
\mathcal{N}(\mathbf{0}, \Sigma_\xi)$

\STATE Form all $N_wN_\xi$ parameter--noise scenario pairs

\STATE $\mathcal{C}_{\mathrm{best}}\gets+\infty$

\FOR{$r=1,\ldots,N_{\mathrm{CEM}}$}

    \FOR{$p=1,\ldots,P$}

        \STATE Sample $\mathbf{U}^{(p)}
\gets \boldsymbol{\mu}_U+
\boldsymbol{\sigma}_U\odot\mathbf{E}^{(p)}$,
$E_{ab}^{(p)}\overset{\mathrm{i.i.d.}}{\sim}\mathcal{N}(0,1)$

        \STATE Project $\mathbf{U}^{(p)}$ onto the admissible
        input set $\mathcal{U}^{\ell}$

\STATE $\mathcal C^{(p)} \gets
\mathcal C_t^{(\ell)}(\mathbf U^{(p)};\mathbf s_t)$
using the shared $N_wN_\xi$ scenarios

    \ENDFOR

    \STATE $\mathcal{E}\gets$ indices of the $P_e$
    lowest-cost candidates

    \STATE $p^\star \gets \arg\min_{p=1,\ldots,P}\mathcal C^{(p)}$
\IF{$\mathcal C^{(p^\star)} < \mathcal C_{\rm best}$}
    \STATE $(\mathcal C_{\rm best},\mathbf U_t^{(\ell)\star})
    \gets(\mathcal C^{(p^\star)},\mathbf U^{(p^\star)})$
\ENDIF

    \STATE $(\bar{\boldsymbol{\mu}}_U,\bar{\boldsymbol{\sigma}}_U)
\gets
\bigl(
\operatorname{Mean}_{p\in\mathcal E}\mathbf U^{(p)},
\operatorname{Std}_{p\in\mathcal E}\mathbf U^{(p)}
\bigr)$

    \STATE
    $\boldsymbol{\mu}_U
    \gets
    \beta\boldsymbol{\mu}_U
    +(1-\beta)\overline{\boldsymbol{\mu}}_U$

    \STATE
    $\boldsymbol{\sigma}_U
    \gets
    \max\!\left\{
    \beta\boldsymbol{\sigma}_U
    +(1-\beta)\overline{\boldsymbol{\sigma}}_U,
    \sigma_{\min}
    \right\}$

\ENDFOR

\STATE
$\boldsymbol{\mu}_U
\gets
[\boldsymbol{\mu}_U(:,2:\ell),
 \boldsymbol{\mu}_U(:,\ell)]$

\STATE
$\boldsymbol{\sigma}_U
\gets
[\boldsymbol{\sigma}_U(:,2:\ell),
 \boldsymbol{\sigma}_U(:,\ell)]$

\RETURN
$\mathbf{U}_t^{(\ell)\star},
\boldsymbol{\mu}_U,\boldsymbol{\sigma}_U$

\end{algorithmic}
\end{algorithm}

\section{Numerical case study: active learning of a noisy bistable system}
\label{sec:numerical_case_study}

\begin{figure*}
    \centering
    \includegraphics[width=\textwidth]{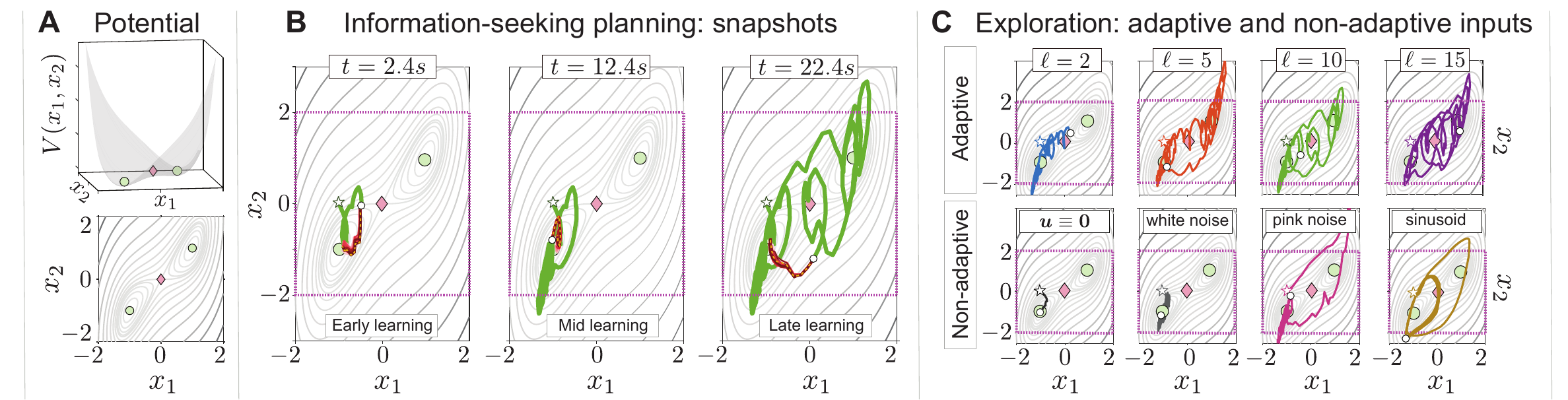}
    \vspace{-0.25cm}
    \caption{
    \begin{small}
    \textbf{Bistable system geometry, non-myopic information-seeking planning, and resulting exploration trajectories. (A)} \textbf{Potential landscape} of the unforced double-well system with $a = k = 1$, showing the two attraction basins and the state-space projection of the state-input region $\mathcal{Z}$ where we aim to learn the dynamics---represented by a dotted purple square in the remaining panels. \textbf{(B)} 
    \textbf{Information-seeking planning snapshots ($\ell = 10$):} Representative non-myopic planning snapshots for the active-learning controller at early, intermediate, and late stages of learning.
    Thin red curves show the state rollouts associated with the top 20 candidate control sequences from the final CEM iteration, ranked according to their objective values (darker curves correspond to higher informativeness),
    while the gold dotted curve denotes the selected plan; the executed state trajectory is shown for context (in green).  The star indicates the initial condition $(-1,0)$, and the current state is represented by a white circle.  \textbf{(C) State-space trajectories generated by adaptive and non-adaptive excitation strategies for the same experiment.} 
    (Top row) Adaptive inputs with planning horizons $ \ell\in\{2,5,10,15\}$, as indicated; (bottom row) the non-adaptive baselines are passive forcing ($\boldsymbol{u} \equiv \boldsymbol{0}$), white noise, pink noise, and sinusoidal forcing. All inputs satisfy the imposed control bounds, and the non-adaptive baselines are power-matched to the active-learning realization corresponding to $\ell = 10$.  
    \end{small}}
    \label{fig:non_myopic_exploration}
\end{figure*}

 We illustrate the proposed active-learning strategy on a controlled stochastic bistable system\footnote{Code available at: \url{https://github.com/jarbelaiz/active-learning-dynamics}}. The autonomous drift is generated by the potential
\begin{equation}
    V(x_1,x_2)
    =
    a\left(\frac{x_1^4}{4}-\frac{x_1^2}{2}\right)
    +
    \frac{k}{2}(x_2-x_1)^2,
    \label{eq:double_well_potential}
\end{equation}
where $a>0$ sets the depth of its two wells and $k>0$ couples the two
state coordinates. The associated deterministic gradient flow,
$\dot{\mathbf{x}}=-\nabla V(\mathbf{x})$, has two stable equilibria at
$(x_1,x_2)=(\pm1,\pm1)$, separated by a saddle at the origin; see
Fig. \ref{fig:non_myopic_exploration}\textbf{A}. When subject to weak stochastic forcing, double-well
gradient systems provide canonical examples of metastable dynamics:
trajectories can remain in one potential well for long periods before
noise induces a transition across the separating barrier. Such
noise-driven transitions arise naturally in overdamped Langevin models
and, in the small-noise regime, are classically described by
Kramers-type barrier-crossing theory \cite{Kramers1940}. This behavior
makes the example useful to test active learning strategies, as passive trajectories may provide highly redundant samples within one 
basin while leaving other regions of state space poorly explored. Accordingly, we consider the controlled stochastic dynamics
\begin{equation}
    d\mathbf{X}(t)
    =
    \left[-\nabla V(\mathbf{X}(t)) + \mathbf{g}(\mathbf{X}(t))u(t)\right]dt
    +
    Q^{1/2} d\mathbf{W}(t),
    \label{eq:bistable_sde}
\end{equation}
where $\mathbf{W}$ is a two-dimensional standard Wiener process,
$Q\succ 0$ is the diffusion covariance, and
$u(t)\in\mathbb{R}$ is a scalar control input; the system is therefore \textit{underactuated}. The state-dependent input
map is
$
    \mathbf{g}(\mathbf{x})
    =[
        0 \;\;\,
        1+\frac{1}{2}\tanh(x_1) ]^\top,
    \label{eq:bistable_input_map}
$
making the effectiveness of the
single actuator vary across state space. The learner reconstructs the controlled flow field
jointly as a function of state and input---neither the
autonomous drift nor the state-dependent input map is supplied.

\paragraph{Numerical setup}
We simulate~\eqref{eq:bistable_sde} using the Euler--Maruyama scheme
with sampling interval $\Delta t=0.1$. Denoting the resulting discrete
states by $\mathbf{x}_t$,
\begin{equation}
    \mathbf{x}_{t+1}
    =
    \mathbf{x}_t
    +
     \mathbf{h}(\mathbf{x}_t, u_t)
    +
    \boldsymbol{\xi}_t,
    \label{eq:bistable_em}
\end{equation}
where $\boldsymbol{\xi}_t \overset{\mathrm{i.i.d.}}{\sim}\mathcal{N}(\mathbf{0}, \Sigma_\xi)$ with $\Sigma_{\xi}:= \Delta t\, Q$ and 
$
    \mathbf{h}(\mathbf{x},u):=-\Delta t\, \nabla V(\mathbf{x}) + \Delta t\, \mathbf{g}(\mathbf{x})u
$
denotes the deterministic one-step state increment. 

We collect $T=250$ samples during learning, corresponding to a total duration of the experiment of $25\,\mathrm{s}$. The learner uses the estimator
$
    \widehat{\mathbf{h}}(\mathbf{x},u)= \Phi(\mathbf{x},u) \mathbf{w}
$---with $d_\phi = 75$ fixed random $\tanh$ features chosen as described in \hyperref[subsec:feature_selec_recipe]{Appendix A}---
together with the Gaussian posterior described in
\S\ref{sec:Bayesian_learning}, initialized with a zero-mean isotropic Gaussian prior. The input is constrained to
$u_t\in[-3,3]$ and we consider planning horizons
$\ell\in\{2,5,10,15\}$. The ROI over which the learner aims to reproduce dynamics is $\mathcal{Z} = [-2,2]^2 \times [-3,3]$, discretized using a  $17\times17\times11$ target grid. At each replanning step, the CEM uses $200$ candidate control sequences, $25$ elites, and
$7$ optimization iterations. Candidate scores are approximated
using $N_w=10$ posterior parameter samples and $N_{\xi}=10$
 process-noise realizations. Each experiment is repeated over $20$ stochastic realizations. 
These seeds determine the realized process noise, the Monte Carlo
scenarios used to evaluate candidate control sequences, and the
stochastic realizations of the open-loop baselines. That is, each seed
corresponds to a new stochastic realization of the same underlying
identification problem. 

\paragraph{Numerical results}
We compare the proposed adaptive strategy against passive excitation
(i.e., $u_t \equiv 0$) and three non-adaptive inputs: white noise, pink noise, and
sinusoidal forcing. All driven baselines satisfy the same input
constraint and are  
matched to the empirical input
power of the corresponding active-learning realization. We evaluate
learning primarily through the average posterior differential entropy
of the estimated controlled flow field over the target grid in
$\mathcal{Z}$, and use the relative root-mean-square error (r-RMSE) over the same grid as an offline validation metric:
$
\text{r-RMSE}_t:= \left[
\frac{1}{N_*}\sum_{i=1}^{N_*}
\frac{\|\widehat{\mathbf h}_t(\mathbf{z}_*^{(i)})-\mathbf h(\mathbf{z}_*^{(i)})\|_2^2}
     {(\|\mathbf h(\mathbf{z}_*^{(i)})\|_2+\epsilon)^2}
\right]^{1/2},
$
where $\epsilon$ is a small constant preventing singular normalization near zero-flow locations.

\begin{figure}
    \centering
\includegraphics[width=\linewidth]{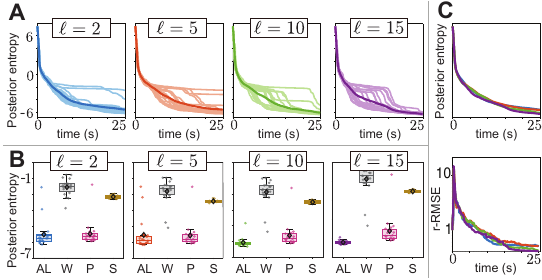}
    \caption{
    \begin{small}
    \textbf{Learning performance across stochastic realizations and comparison with non-adaptive excitation strategies. (A)} Time evolution of the grid-averaged posterior entropy for planning horizons $\ell\in\{2,5,10,15\}$ over $20$ stochastic realizations. Each thin curve corresponds to one realization, while the thick curve denotes the pointwise mean across seeds; the same axes are used for all horizons. Longer planning horizons generally yield greater uncertainty reduction and lower terminal posterior entropy, although individual learning trajectories vary across realizations. \textbf{(B)} Distribution of the final grid-averaged posterior entropy for the active-learning controller (AL) and the power-matched non-adaptive baselines: white noise (W), pink noise (P), and sinusoidal forcing (S). Dots denote individual seeds, diamonds sample means, horizontal lines in each box the corresponding median, and the box spans the interquartile range.
    Each panel corresponds to the planning horizon $\ell$ indicated above it.  \textbf{(C)}  Mean learning trajectories across seeds for all planning horizons, using the same horizon color coding as in (A), shown for grid-averaged posterior entropy (top) and relative RMSE (bottom). 
    \end{small}
    }
    \label{fig:learning_performance}
\end{figure}

Fig. \ref{fig:non_myopic_exploration}\textbf{B} illustrates the non-myopic information-seeking mechanism underlying the
closed-loop.
The resulting trajectories in Fig. \ref{fig:non_myopic_exploration}\textbf{C} show
qualitatively different state-space coverage across planning horizons
and excitation strategies, with longer-horizon policies more readily
producing excursions between attraction basins. Fig. \ref{fig:learning_performance}\textbf{A} shows the evolution of the posterior entropy across
stochastic realizations for different planning horizons $\ell$: longer planning horizons generally accelerate learning and
reduce the prevalence of realizations that remain at high posterior
uncertainty at the end of the experiment. Fig. \ref{fig:learning_performance}\textbf{B} compares
terminal performance with the non-adaptive baselines. The active
controller substantially outperforms spectrally white excitation,
while pink noise is the strongest open-loop competitor in this bistable system---as the long temporal correlations of pink noise can generate sustained forcing and induce large excursions across the
bistable landscape.  However, this favorable temporal structure is
specified \textit{a priori}, whereas the \textit{proposed controller adapts online to the evolving posterior uncertainty and observed
state in an automated manner}. The same qualitative trend is reflected by the offline r-RMSE
results in Fig. \ref{fig:learning_performance}\textbf{C}, indicating that reductions in posterior entropy are accompanied by improved reconstruction of the
underlying controlled flow field.

\section{Conclusion }
\label{sec:conclusion}

We introduced a prediction-oriented Bayesian active-learning framework
for online identification of stochastic nonlinear dynamical systems over
a prescribed region of state--input space. Informativeness is quantified
 in terms of the reconstructed dynamics through a mean marginal
mutual-information criterion. We cast adaptive input design as a
sequential decision problem over an information state comprising the
physical state and the current Bayesian posterior, and approximate the
resulting causal design problem using non-myopic receding-horizon
planning. For dynamics models that are linear in the unknown parameters,
the predictive information gain can be evaluated analytically, while
the remaining expectation over uncertain future trajectories is
approximated using a scenario-based sample average and optimized with
CEM. Numerical experiments on a noisy bistable system show that the
resulting closed-loop strategy efficiently reduces predictive uncertainty
and reconstruction error over the target region.

\section*{
Acknowledgment
}

JA acknowledges useful discussions with Memming Park and Hyungju Jeon (Champalimaud Centre for the Unknown), and with Prof. H. J. Kappen (Radboud University).

\section*{Appendix}

\subsection*{Appendix A: Random-feature construction}
\label{subsec:feature_selec_recipe}

We construct the fixed random features over the state--input ROI
$
    \mathcal{Z}
    =
    \prod_{p=1}^{n+m}
    [z_p^{\min},z_p^{\max}]
    \subset\mathbb{R}^{n+m}.
$
Let
$
    L_p:=z_p^{\max}-z_p^{\min}
$
and
$
    \overline{L}:=\frac{1}{n+m}
    \sum_{p=1}^{n+m}L_p
$
denote the coordinate widths of $\mathcal{Z}$ and their mean,
respectively. For each feature
$
    \phi_i(\mathbf{z})
    = \tanh(\mathbf{j}_i^\top\mathbf{z}+b_i),\;  i=1,\ldots,d_\phi,
$
we draw
$
    \mathbf{g}_i\sim\mathcal{N}(\mathbf{0},\mathbb{I}_{n+m}),
    \;
    \mathbf{v}_i
    :=
    \frac{\mathbf{g}_i}{\|\mathbf{g}_i\|_2},
$
and set
$
    \mathbf{j}_i
    :=
    \frac{\rho}{\overline{L}}\mathbf{v}_i,
$
where $\rho>0$ controls the characteristic variation of the features
over the ROI. We then sample a center
$\mathbf{z}_i^\circ$ uniformly from $\mathcal{Z}$ and choose
$
    b_i:=-\mathbf{j}_i^\top\mathbf{z}_i^\circ.
$
Hence,
$
    \phi_i(\mathbf{z})
    =
    \tanh\!\left(
        \mathbf{j}_i^\top(\mathbf{z}-\mathbf{z}_i^\circ)
    \right),
$
so the zero-level hyperplane of each feature passes through a randomly
sampled point in the ROI. 

Normalizing the random directions removes
variability in their norms, 
while scaling their norms inversely with the characteristic width of
 $\mathcal{Z}$ and distributing their transition regions
throughout the ROI helps avoid widespread feature saturation. The
sampled feature parameters $(J,\mathbf{b})$ are held fixed throughout
learning; in the numerical experiments we use $\rho=1$.

\vspace{-0.1cm}
\subsection*{Appendix B: Posterior over $\vw$}
\label{subsec:app_posterior_over_w}

We derive the \textit{batch posterior update} in~\eqref{eq:batch_update}.
Given the dataset
$\mathcal{D}_t:=\{\mathbf{x}_{0:t},\mathbf{u}_{0:t-1}\}$,
a sequence of inputs $\mathbf{u}_{t:t+\ell-1}$, and the resulting
state observations $\mathbf{x}_{t+1:t+\ell}$, Bayes' rule gives
$
p\!\left(
    \mathbf{w}
    \mid
    \mathbf{x}_{t+1:t+\ell},
    \mathbf{u}_{t:t+\ell-1},
    \mathcal{D}_t
\right) \propto
p\!\left(
    \mathbf{x}_{t+1:t+\ell}
    \mid
    \mathbf{w},
    \mathbf{u}_{t:t+\ell-1},
    \mathcal{D}_t
\right)
p(\mathbf{w}\mid\mathcal{D}_t).
$
Conditioned on $\mathbf{w}$, the dynamics in~\eqref{eq:parametric_dynamics}
are controlled Markov and the process disturbances are independent;
hence,
$
p\!\left(
    \mathbf{x}_{t+1:t+\ell}
    \mid
    \mathbf{w},
    \mathbf{u}_{t:t+\ell-1},
    \mathcal{D}_t
\right)
=
\prod_{\tau=t+1}^{t+\ell}
p\!\left(
    \mathbf{x}_{\tau}
    \mid
    \mathbf{x}_{\tau-1},
    \mathbf{u}_{\tau-1},
    \mathbf{w}
\right).
$
Defining
$\mathbf{y}_{\tau}:=\mathbf{x}_{\tau}-\mathbf{x}_{\tau-1}$
and
$\Phi_{\tau-1}:=
\Phi(\mathbf{x}_{\tau-1},\mathbf{u}_{\tau-1})$,
each likelihood factor is equivalently
$
    p(\mathbf{y}_{\tau}\, | \, \mathbf{w})
    =
    \mathcal{N}\!\left(
        \Phi_{\tau-1}\mathbf{w},
        \Sigma_{\xi}
    \right).
$
Combining these likelihood terms with
$p(\mathbf{w}\, | \, \mathcal{D}_t)
=
\mathcal{N}(\mathbf{m}_t, \Omega_t)$
and dropping terms independent of $\mathbf{w}$ yields
$
\log p(\mathbf{w} \,| \, \mathbf{x}_{t+1:t+\ell},
    \mathbf{u}_{t:t+\ell-1},
    \mathcal{D}_t )
= -\frac{1}{2}
\sum_{\tau=t+1}^{t+\ell}
\left(
    \mathbf{y}_{\tau}-\Phi_{\tau-1}\mathbf{w}
\right)^{\!\top}
\Sigma_{\xi}^{-1}
\left(
    \mathbf{y}_{\tau}-\Phi_{\tau-1}\mathbf{w}
\right)
-
\frac{1}{2}
\left(
    \mathbf{w}-\mathbf{m}_t
\right)^{\!\top}
\Omega_t^{-1}
\left(
    \mathbf{w}-\mathbf{m}_t
\right)
+\mathrm{const}.
$
Collecting the quadratic and linear terms in $\mathbf{w}$ gives the
Gaussian posterior with mean and covariance in
\eqref{eq:batch_update}. Setting $\ell=1$ and applying the matrix
inversion lemma (Woodbury identity) yields the \textit{recursive update}
in~\eqref{eq:recursive_update}.

\bibliographystyle{IEEEtran}
\bibliography{refs}

\end{document}